\documentclass{desyproc}

\begin{document}
\title{\emph{a}KWISP: investigating short-distance interactions at sub-micron scales}

\author{{\slshape G. Cantatore$^1$, V. Anastassopoulos$^{2}$, S. Cetin$^{3}$, H. Fischer$^{4}$, W. Funk$^{5}$, A Gardikiotis$^{2}$, D.H.H. Hoffmann$^{6}$, M. Karuza$^{7,1}$, Y.K. Semertzidis$^{8}$, D. Vitali$^{9}$, K. Zioutas$^{2,5}$}\\[1ex]
$^1$University and INFN Trieste, Trieste, Italy\\
$^2$University of Patras, Patras, Greece\\
$^3$Bilgi University, Istanbul, Turkey\\
$^4$University of Freiburg, Freiburg, Germany\\
$^5$CERN, Geneva, Switzerland\\
$^6$TU Darmstadt, Darmstadt, Germany\\
$^7$University of Rijeka, Rijeka, Croatia\\
$^8$CAPP, Daejeon, Korea\\
$^9$University of Camerino, Camerino, Italy}

\contribID{Cantatore_Giovanni}

\confID{16884}  
\desyproc{DESY-PROC-2017-02}
\acronym{Patras 2017} 
\doi  

\maketitle

\begin{abstract}
The sub-micron range in the field of short distance interactions has yet to be opened to experimental investigation, and may well hold the key to understanding al least part of the dark matter puzzle. The \emph{a}KWISP (\emph{advanced}-KWISP) project introduces the novel Double Membrane Interaction Monitor (DMIM), a combined source-sensing device where interaction distances can be as short as 100 nm or even 10 nm, much below the $\approx 1-10$ $\mu$m distance which is the lower limit encountered by current experimental efforts. \emph{a}KWISP builds on the technology and the results obtained with the KWISP opto-mechanical force sensor now searching at CAST for the direct coupling to matter of solar chameleons. It will reach the ultimate quantum-limited sensitivity by exploiting an array of technologies, including operation at milli-Kelvin temperatures. Recent suggestions point at short-distance interactions studies as intriguing possibilities for the detection of axions and of new physical phenomena.
\end{abstract}

\section{Introduction}

Sensitive measurements on interactions at short separation distances between macroscopic bodies provide a window on physics beyond the Standard Model. In this field of study, interest focuses on Casimir-type interactions, including the topological Casimir effect which might lead to the detection of axions~\cite{Zhitnitsky:2017}, and on possible deviations from the standard gravitational interaction, modelled by a Yukawa-type potential representing the exchange of force carrier scalar particles~\cite{Tkachev:1982,Moody:1984}. These particles might be, for example, Dark Matter or Dark Energy candidates such as axions, moduli and chameleons, portals to extra-dimensions or dilatons. Each single one of these themes opens a view beyond the Standard Model.

A key parameter is the distance scale at which the interaction is probed. Current experimental efforts reach distances of the order of 1-10 $\mu$m ~\cite{Chen:2016} . Recently, even collider experiments such ATLAS at LHC reported probing extra-dimensions down to a distance of 11 $\mu$m~\cite{ATLAS_note}.

The original idea of \emph{a}KWISP is building a novel opto-mechanical device consisting of two micro-membranes set a short distance apart, down to a few tens of nm. One of the two membranes (the ``source'' body) is excited in a controlled way by the radiation pressure of an amplitude-modulated laser beam. The displacements of the other membrane (the ``sensing'' body), possibly due to interactions with the source body, are then monitored by a Fabry-Perot interferometer at resonance with a second laser beam. Displacement sensitivities as low as $10^{-15} \ \mbox{m}/\sqrt{\mbox{Hz}}$ can be reached at room temperature~\cite{Karuza:2016} thanks to the combination of two large quality factors, the finesse of the Fabry-Perot optical resonator $(\approx10^{5})$ and the figure of merit of the membrane mechanical resonator $(\approx10^{5})$.

\emph{a}KWISP enters the short-distance interaction field with the novel Double Membrane Interaction Monitor (DMIM) concept where interaction distances can be as short as 10 nm and  plans to reach the ultimate quantum-limited sensitivity by achieving milli-Kelvin equivalent membrane temperatures with a combination of cryogenic and optical cooling~\cite{Karuza:2012}.

\begin{figure}[hb]
\centerline{\includegraphics[width=0.5\textwidth]{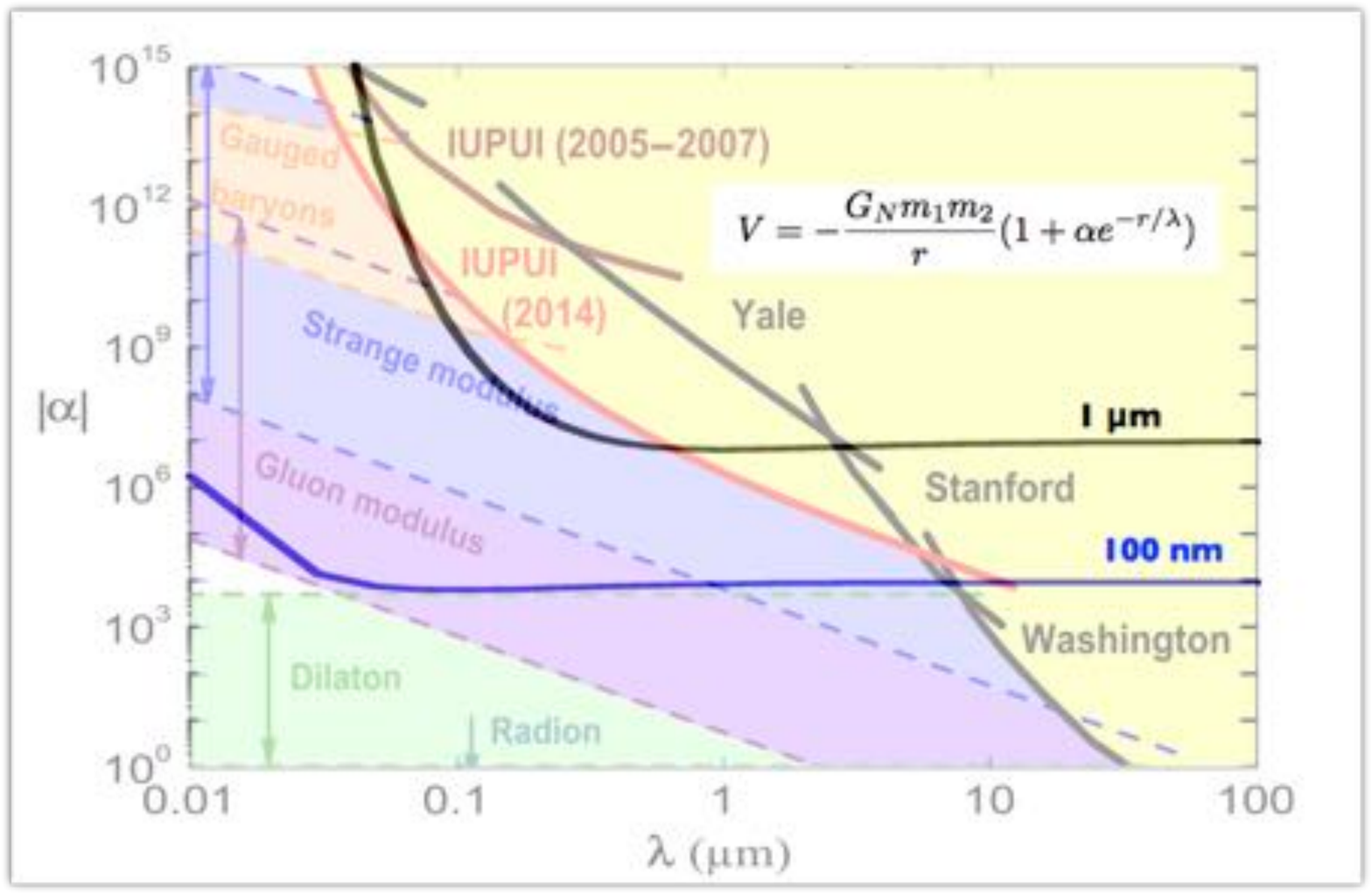}}
\caption{Projected \emph{a}KWISP coverage in the parameter space of a Yukawa-type interaction between two macroscopic bodies. $\alpha$ and $\lambda$ represent the interaction strength and range, respectively. The black curve refers to a 1 $\mu$m separation distance between the two DMIM membranes, while the blue curve refers to a 100 nm one (see also text). The background plot is taken from~\cite{Chen:2016}.}\label{Fig:PS}
\end{figure}

Figure~\ref{Fig:PS} shows an example of projected coverage by \emph{a}KWISP in the parameter space of Yukawa-type corrections to the standard gravitational potential between two macroscopic bodies. $\alpha$ and $\lambda$ represent the interaction strength and range, respectively. This coverage is obtained when operating \emph{a}KWISP at a temperature of 3~mK with thermally limited sensitivity and $10^5$~s integration time, corresponding to the detection of a force as feeble as $10^{-20}$~N.
 
 \section{The \emph{a}KWISP experimental setup}
 
\begin{figure}[hb]
\centerline{\includegraphics[width=0.5\textwidth]{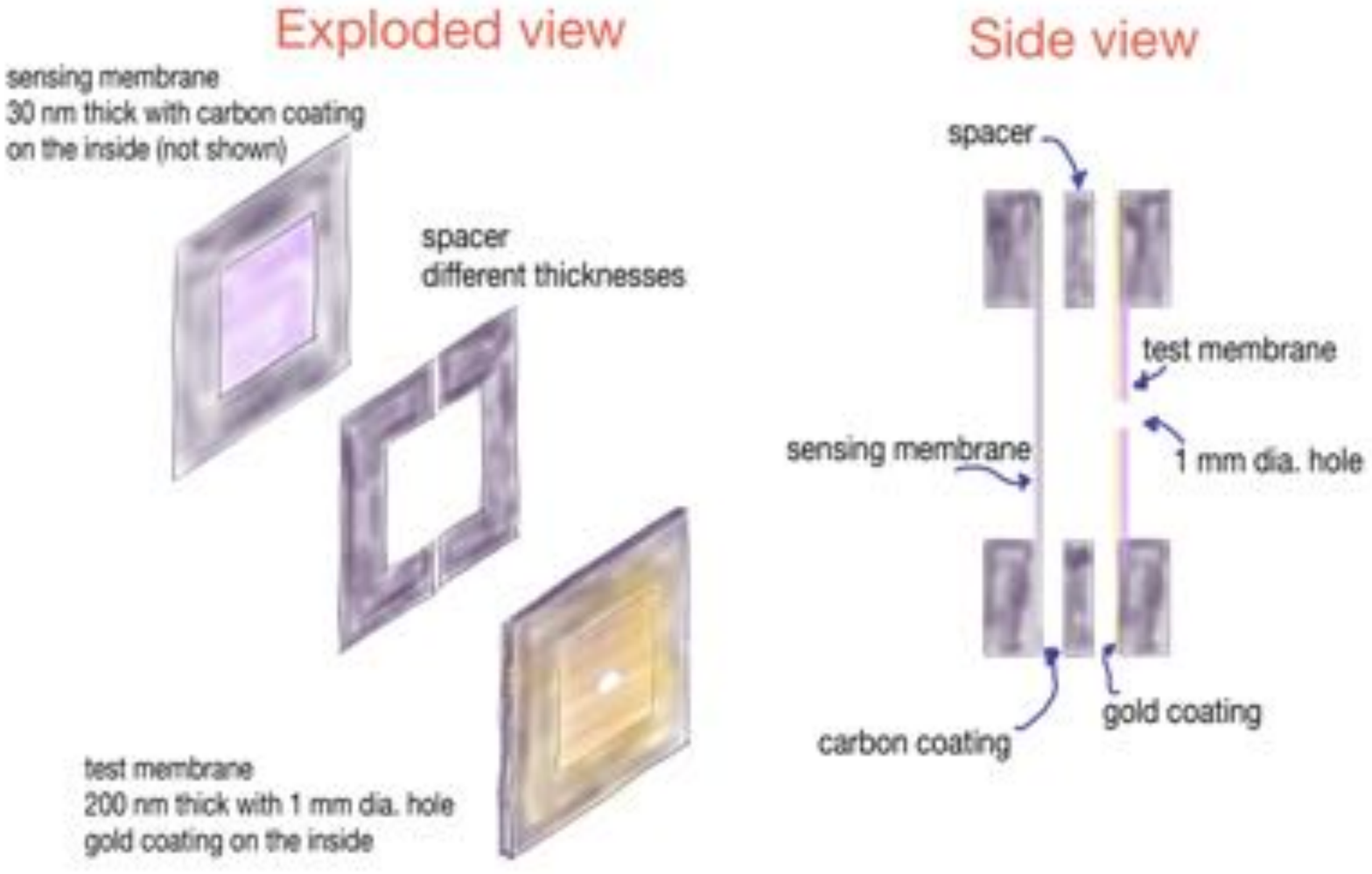}}
\caption{Sketch of the Double Membrane Interaction Monitor (DMIM) (see text).}\label{Fig:DMIM}
\end{figure}

The \emph{a}KWISP concept is based on the Double Membrane Interaction Monitor (DMIM) device. It consists of two thin (thickness $< 200$ nm) and taut $\mbox{Si}_{3}\mbox{N}_{4}$ membranes set parallel to each other at a distance which can be as short as 10 nm. Typical membrane dimensions are $5\cdot5 \ \mbox{mm}^{2}$. One of the two membranes is coated with an Al layer in order to be able to excite its oscillations in a controlled way by reflecting a laser beam off it, and in addition it has a 1 mm diameter hole at its center to allow free passage of a second laser beam which monitors the displacements of the second membrane. Both membranes have a 3 nm thick carbon conductive coating to allow setting them at a common electrical potential  in order to reduce the effects of static charges. 
Figure~\ref{Fig:DMIM} shows a sketch of the DMIM device we have invented for \emph{a}KWISP. A prototype set of suitable membranes for a DMIM has already been produced by Norcada Inc., Canada, and will be tested in the INFN Trieste optics laboratory.

In the \emph{a}KWISP experimental apparatus, the DMIM is  placed inside an optical Fabry-Perot interferometer kept at resonance, by means of an electro-optic active feedback~\cite{Cantatore:1995}, with a 1064 nm ``sensing'' laser beam. The Fabry-Perot is mounted inside a vacuum chamber to avoid noise due to the Brownian motion of air.

\begin{figure}[hb]
\centerline{\includegraphics[width=0.5\textwidth]{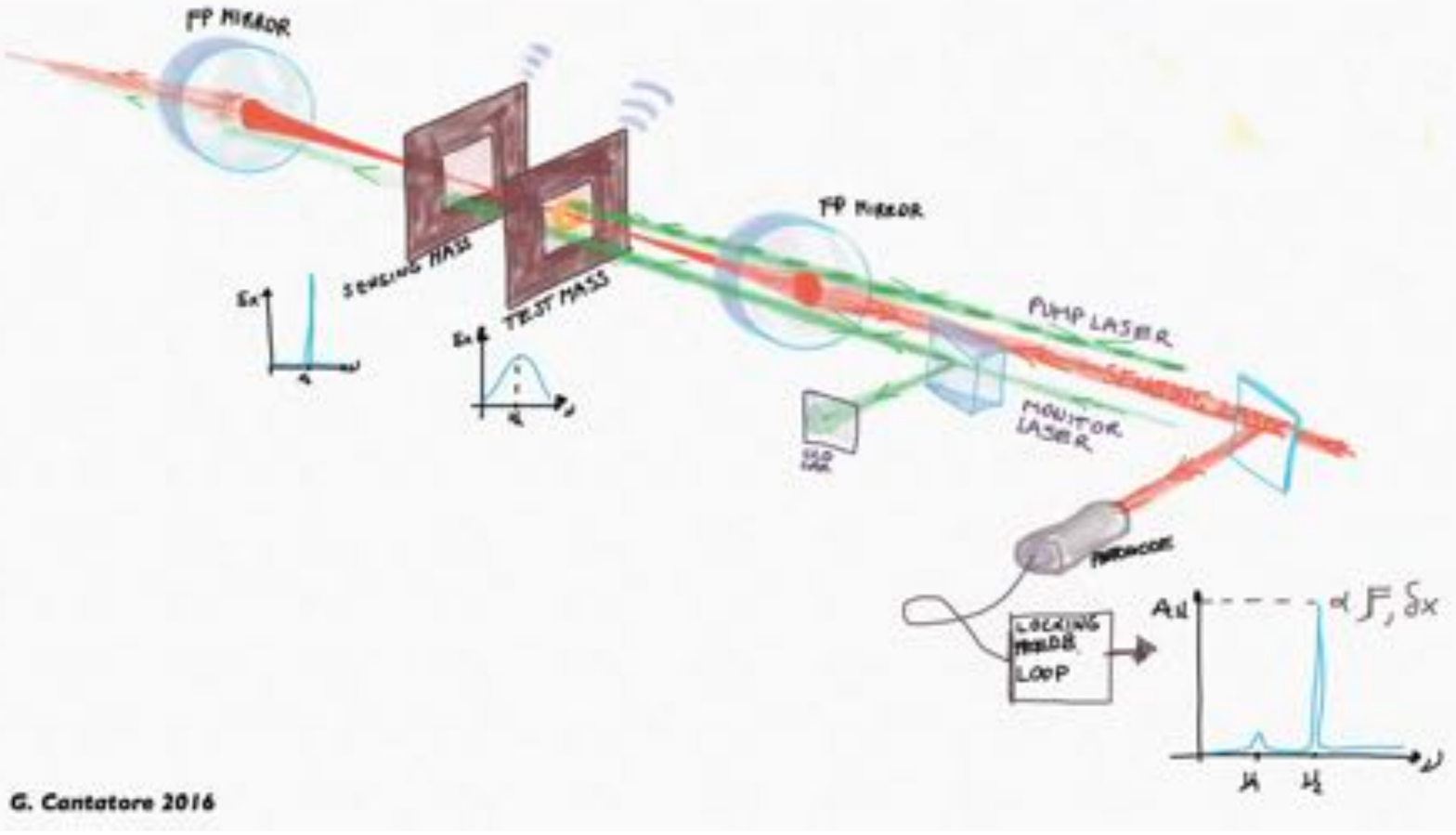}}
\caption{Conceptual schematic of a possible \emph{a}KWISP experimental setup (see text).}\label{Fig:Setup}
\end{figure}

Thanks to the hole in the source membrane, the sensing beam measures only the displacements of the sensing membrane: these modify the frequency of the standing optical wave resonating within the Fabry-Perot causing a detuning, which is then detected by the electro-optic feedback loop. This sensing scheme has already been demonstrated in~\cite{Karuza:2016} where a single-membrane sensor has achieved a force sensitivity of $1.5\cdot10^{-14} \ \mbox{N}/\sqrt{\mbox{Hz}}$ ($7.5\cdot10^{-16} \ \mbox{m}/\sqrt{\mbox{Hz}}$  in terms of displacement) at room temperature. An opto-mechanical particle detector based on this principle, called KWISP, is now searching at CAST for the direct coupling to matter of solar chameleons~\cite{KWISP:2014}. A second laser beam (``pump'' beam), operating at a different wavelength (532 nm in our case) where the mirrors of the Fabry-Perot resonator are practically transparent, is used to excite oscillations in the source membrane by reflecting it off the Al coating. The frequency and amplitude of these oscillations can be controlled by a suitable amplitude modulation impressed on the pump beam. Figure~\ref{Fig:Setup} shows a schematic sketch of a possible setup for \emph{a}KWISP.

The ultimate sensitivity reachable with this class of opto-mechanical devices depends on temperature~\cite{Lamoreaux:2008}. In order to reach equivalent temperatures in the milli-Kelvin range the DMIM could be inserted in the payload of a suitable cryostat/refrigerator and further cooled by means of optical cooling techniques~\cite{Karuza:2012}.

\section{Perspectives}
Highly sensitive measurements on short-distance interactions may uncover a host of hitherto unobserved physical processes, opening a window beyond the Standard Model. Among these, axions, moduli and chameleon particles, as well as portals towards extra dimensions and perhaps dilatons, may play a role. Casimir-type forces are also an open field of study when interaction distances range below 1 $\mu$m. Recently, the ``topological'' Casimir effect, which has never been observed, has been proposed as a tool to search for axions~\cite{Zhitnitsky:2017}.
The key parameter in all these investigations is the distance between the two interacting bodies, which must be less than 1 $\mu$m in order to open access to unexplored regions in parameter space (see Figure~\ref{Fig:PS}). The DMIM proposed here has the potential of lowering this distance to reach the 100 nm or perhaps even the 10 nm range. Using this device, the \emph{advanced}-KWISP proposal presents a novel scheme to study sub-nuclear scale phenomena using a high sensitivity table-top opto-mechanical sensor based on precision technologies, such as micro-membranes and interferometric sensing coupled to cryogenic and optical cooling




\begin{footnotesize}

\end{footnotesize}


\end{document}